\documentclass[12pt]{iopart}

\expandafter\let\csname equation*\endcsname\relax
\expandafter\let\csname endequation*\endcsname\relax

\usepackage{booktabs,multirow,array,multirow}
\usepackage{amssymb}
\usepackage{float}
\usepackage{setspace}
\usepackage{amsmath}
\usepackage{textcomp}
\usepackage{url}
\usepackage[colorlinks=true, allcolors=blue]{hyperref}
\usepackage{color}
\usepackage{MnSymbol}
\usepackage{lipsum}
\usepackage{amsfonts}
\usepackage{graphicx}
\usepackage{dcolumn}
\usepackage{tensor}
\usepackage{bm}
\usepackage{hyperref}
\usepackage{soul}
\usepackage{xcolor}
\usepackage{svg}
\newcommand{\orcid}[1]{\href{https://orcid.org/#1}{\includesvg[width=10pt]{orcid.svg}}}

\usepackage[normalem]{ulem}
\newcommand{\stkout}[1]{\ifmmode\text{\sout{\ensuremath{#1}}}\else\sout{#1}\fi}

\definecolor{Red}{rgb}{0.9,0,0}

\begin{document}

\title{On the connection between Lenz's law and relativity}


\author{Thales B. S. F. Rodrigues  \& B. F. Rizzuti}

\address{Instituto de Ciências Exatas, Departamento de Física, Universidade Federal de Juiz de Fora, Juiz de Fora, MG, Brazil}
\ead{thalesfonseca@ice.ufjf.br \& brunorizzuti@ice.ufjf.br}
\vspace{10pt}

\begin{abstract}
In this work, we demonstrate explicitly the unified nature of electric and magnetic fields, from  the principles of special relativity and Lorentz transformations of the electromagnetic field tensor. Using an operational approach we construct the tensor and its corresponding transformation law, based on the principle of relativity. Our work helps to elucidate concepts of advanced courses on electromagnetism for primary-level learners and shows an alternative path to derive the Lenz's law based  on the connection between relativity arguments and a standard electromagnetism experiment.
\end{abstract}

%
%
%
%
%

\section{Introduction}
\label{sec1}

 In most introductory courses on electromagnetism, undergraduate students accept the idea that the electric and magnetic fields are independent entities. One of the reasons why it happens could be related to the duration of the program: when it is time to unity them, the semester is over. Usually, in textbooks used in these courses, first, the student learns about the electric field separately before the magnetic field. It's only after that the Maxwell's equations and their applications are introduced, suggesting a liaison with electric and magnetic phenomena on either side \cite{young, moyses3}. Yet the connection between these allegedly different topics is not explored.

On the other hand, it is only in advanced courses and textbooks on electromagnetism \cite{jackson_classical_1999} that the students truly learn about the nature of electric and magnetic fields and their behavior as a unified entity. We need to reconsider the current approach to teaching these concepts and explore the possibility of introducing such formulations in primary-level courses (i.e., courses that teach elementary concepts of specific topics in physics at the college level).

In \cite{RL_QM}, an approach was presented that establishes the connection between the special theory of relativity and introductory courses in quantum mechanics. This approach demonstrates that it is feasible, from a pedagogical standpoint, to introduce these advanced concepts to primary-level learners. Similar approaches have been observed in the literature regarding the teaching of electromagnetic field concepts \cite{Sobouti_2015,Heras_2020,FeynmanVol2,ramos_braga_silva_lima_holanda_2017, teixeira_de_moraes_instructional_2022}. As in \cite{König_2021}, for instance, by demonstrating that Lorentz force is not Galilean invariant, the author highlighted the inherent need for a new transformation connecting different inertial systems.This was achieved by employing an infinitesimal Lorentz transformation approach, ultimately ensuring the invariance of the Lorentz force.

However, up to our knowledge, it is not explicitly addressed in other works in literature how the Lorentz transformations of the electromagnetism field tensor $\mathcal{F}^\mu{}_\nu$ play a crucial role in unifying the electric and magnetic fields, giving birth to the induction/Lenz's law. On one hand, works in literature either focus their approach on demonstrating the lack of covariance of Maxwell equations and the Lorentz force under Galilean transformations, or on alternative approaches to evince electromagnetism covariance under Lorentz transformation. On the other hand, what sets our work apart is the primary emphasis on the relativistic origin of Lenz's law, based on the transformation rule of the tensor $\mathcal{F}^\mu{}_\nu$.

In this work, we aim to propose a didactic approach that demonstrates the unified nature of the electric and magnetic fields to primary electromagnetism students at undergraduate level, anchoring in the principles of special relativity \cite{bernbook} and the coordinate transformation of 
$\mathcal{F}^\mu{}_\nu$. In what follows, first, we introduce the electromagnetic field tensor $\mathcal{F}^\mu{}_\nu$ in section \ref{sec2}, constructed from experimental assumptions already presented in past works \cite{Alves_Rizzuti_Gonçalves_2020}. Then, in section \ref{sec3} we elucidate the transformation rule of the tensor $\mathcal{F}^\mu{}_\nu$, based on Lorentz transformations \cite{rocha_transformacoes_2013}. In section \ref{sec4} and \ref{sec5}, respectively, is showed a recap to the induction law using geometrical and experimental considerations, and, we demonstrate with a simple experiment, the connection between the induction law and the special relativity. In particular, we show how the Lenz's law can be found invoking the principle of relativity. Finally, the conclusions of the work are given in section \ref{sec6}.

Our notation throughout the text will be the following. Greek letters $\alpha$, $\beta$, $\mu$, $\nu$, etc run values from $0$ to $3$, representing 
space-time quantities whereas Latin letters $i$, $j$, $k$, etc assume values $1$, $2$, $3$, meaning spatial quantities. As usual, Einstein summation notation is adopted.

\section{Construction of the electromagnetic field tensor}
\label{sec2}
Tensors are geometric objects that are fundamental in physics due to the fact that they put forward a coordinate-independent description of physical laws \cite{shapiro_primer_2019}.
In this section, we will show the steps for constructing the electromagnetic field tensor $\mathcal{F}^\mu{}_\nu$, which is key for our further considerations. Our staring point will be the Lorentz force a charged particle fells when set in a spatial region filled with electromagnetic field. 
\begin{equation}\label{1.0}
    \Vec{F} = \Vec{F}_E + \Vec{F}_M = q(\Vec{E}  + \vec{v} \times \Vec{B}) .
\end{equation}
As usual, $q$ denotes the particle charge, while $\Vec{E}$ and $\Vec{B}$ stand for the electric and magnetic sectors of the field. 

First, we can construct the magnetic field contribution to the force that acts on charged particles, based on experimental results \cite{Alves_Rizzuti_Gonçalves_2020}, namely, \textit{(i)} the magnetic force $\Vec{F}_M$ is perpendicular to the velocity and depends linearly on it and \textit{(ii)} is also linearly proportional to the charge $q$ of a particle beam. In summary, we can write, owing to the aforementioned empirical data, 
\begin{equation}\label{1.01}
    \Vec{F}_M = q\mathcal{M}(\vec{v}),
\end{equation}
where the object $\mathcal{M}$ is a linear operator that acts on the velocity $\vec{v}$
\begin{equation}
    \mathcal{M}:  \mathbb{V} \rightarrow \mathbb{F}.
\end{equation}
$\mathbb{V}$ is the linear space of velocities and $\mathbb{F}$ the space of force vectors. Abstract vectors (elements of say, $\mathbb{V}$ or $\mathbb{F}$) as well as operators gain a concrete structure when we present them by fixing an ordered basis on the corresponding spaces. For now on, we fix an orthonormal ordered basis $\mathcal{B} = \{ \hat{e}_i \}$, $i=1,2,3$, that, with adjusted coefficients, it is possible to generate any vector with linear combinations. For example, $\Vec{v} = v^i \hat{e}_i$. That way, the vector $\Vec{v}$ has a unique expansion with components $v^i$ to the basis $\mathcal{B}$ and is represented by the column
\begin{equation}
    \Vec{v} \leftrightarrow \begin{pmatrix}
        v^1 \\
        v^2 \\
        v^3
    \end{pmatrix}.
\end{equation}

Linear operators, in turn, are in a one-to-one correspondence with matrices, whose order depends on the dimensions of domain and image of the operator. This is a central result in representing objects in Linear Algebra \cite{hoffman_kunze}. In our particular case, both $\mathbb{V}$ and $\mathbb{F}$ are three dimensional. Thus, $\mathcal{M}$ is represented by a $3\times 3-$matrix. Therefore, we can write \eqref{1.01} as 
\begin{equation}\label{2.4}
     \begin{pmatrix}
        F^1 \\
        F^2 \\
        F^3
     \end{pmatrix}    
        = q 
        \begin{pmatrix}
            \mathcal{M}^1{}_1 & \mathcal{M}^1{}_2 &\mathcal{M}^1{}_3 \\
            \mathcal{M}^2{}_1 & \mathcal{M}^2{}_2 &\mathcal{M}^2{}_3\\
            \mathcal{M}^3{}_1 & \mathcal{M}^3{}_2 &\mathcal{M}^3{}_3
        \end{pmatrix}        
     \begin{pmatrix}
        v^1 \\
        v^2 \\
        v^3
     \end{pmatrix}.   
\end{equation}
By fixing a specific initial direction of the force vector $\Vec{F}_M$ and comparing the components of $\mathcal{M}$ that vanish using the experimental considerations (i) and (ii) described above, we can conclude - see the details in \cite{Alves_Rizzuti_Gonçalves_2020} - that $\mathcal{M}$ is represented by an anti-symmetric matrix, whose elements we denote $B_i = \frac{1}{2}\varepsilon_{ijk} 
\mathcal{M}{^j{}_k}$. As usual, $\varepsilon_{ijk}$ is the Levi-Civita symbol.

With this notation, the expression \eqref{2.4} can be written in a more sympathetic form as 
\begin{equation}\label{seven}
    \Vec{F}_M = q\vec{v}\times \Vec{B}.
\end{equation}
For a detailed description of the last steps, we suggest interested readers to check the construction in the previous study \cite{Alves_Rizzuti_Gonçalves_2020}. The vector notation from \eqref{2.4} to \eqref{seven} is not mere notation. Actually, under rotations, the coefficients $B_i$ transform as components of a vector, justifying the nomenclature.

With the magnetic field contribution at hand, the second step is to introduce the electric field contribution of the object $\mathcal{F}$, which is responsible for yielding the electric force $\Vec{F}_E$. It can be achieved with the help of $4-$vectors. To differentiate them from $3-$vectors, say $\Vec{v}$, $\Vec{F}$, etc, the notation $\overrightharpoon{A}$ will be used henceforth. We start with the $4-$velocity $\overrightharpoon{V}$, whose components, after fixing coordinates (or a basis), are \cite{ugarov}  
\begin{equation} \label{1.8}
    \left (\overrightharpoon{V} \right )^\mu \equiv V^\mu = \gamma
    \begin{pmatrix}
        c \\
        v^1\\
        v^2\\
        v^3
    \end{pmatrix},
\end{equation}
where $\gamma$ is the Lorentz factor
\begin{equation}
     \gamma = \frac{1}{\sqrt{1 - \frac{\Vec{v}^2}{c^2}}},
\end{equation}
as well as $c$ is the speed of light.
The $4-$force $\overrightharpoon{f}$ components also gain a $gamma-$factor in front of it \cite{bernbook}
\begin{equation} \label{1.9}
    \left ( \overrightharpoon{f} \right )^\mu = f^\mu = \gamma
    \begin{pmatrix}
        F^0\\
        F^1\\
        F^2\\
        F^3
    \end{pmatrix}.
\end{equation}

At this point, we will demand that the electromagnetic field is a linear operator acting on the charged particle $4-$velocity, resulting in the $4-$force, in resemblance to the expression we obtained in \eqref{2.4}. We denote it as $\mathcal{F}$ and due to the dimension of the underlying spaces are working on, it will be represented by a $4\times4-$matrix, with components $\mathcal{F}^\mu{}_\nu$. If we define the components of $\mathcal{F}$ as
\begin{equation}\label{fmunu}
    \mathcal{F}^\mu{}_\nu = \begin{pmatrix} 
       0 & E_1 c^{-1}  & E_2 c^{-1} & E_3 c^{-1} \\
       E_1 c^{-1} & 0 & B_3 & -B_2 \\ 
       E_2 c^{-1} &-B_3 & 0 & B_1 \\ 
       E_3 c^{-1} & B_2 &-B_1 & 0
    \end{pmatrix}, 
\end{equation}
then, a direct calculation shows that the spatial sector of the compact form
\begin{equation}
    f^\mu = q \mathcal{F}^\mu{}_\nu V^\nu
\end{equation}
yields \eqref{1.0}, as expected. It is also possible to relate $f^0$ with the power the particle gains due to its interaction with electromagnetic field. This details are not relevant to our discussions and will be omitted here.  

In conclusion, we can see that the electromagnetic field tensor $\mathcal{F}$ is simply a linear operator  which acts on $4-$velocity vectors resulting in $4-$force vectors. \eqref{fmunu} has also a unifying character, in the sense that both $\Vec{E}$ and $\Vec{B}$ are but a manifestation of the electromagnetic field when one fixes a particular basis to represent it. In the next section we will see how it behaves when there is a change in frames of reference.

\section{The transformation rule of $\mathcal{F}^\mu{}_\nu$} \label{sec3}

After the construction presented in the last section, we shall introduce the transformation rule of the electromagnetic field tensor $\mathcal{F}^\mu{}_\nu$, which will be crucial to the whole discussion of this work. 

The coordinate transformation $x\mapsto x'$ of a reference frame to another one with relative velocity $\vec{v} = v \hat{e}_1$ is given by a Lorentz boost $\Lambda$. We will focus on the $x^1-$direction for specific reasons that will be elucidated in the subsequent sections of our work. It is given by
\begin{equation}\label{lorentz-transformation}
    x^\mu \rightarrow x'^\mu = \Lambda^\mu{}_\nu x^\nu. 
\end{equation}
In matrix notation, it reads
\begin{equation} \label{3.17}
    \begin{pmatrix}
         x'^0\\
         x'^1\\
         x'^2\\
         x'^3
    \end{pmatrix}
    = 
    \begin{pmatrix} 
       \gamma & -\beta \gamma  & 0   & 0  \\
       -\beta \gamma & \gamma  & 0   & 0 \\ 
         0 &  0 & 1 & 0 \\ 
         0 &  0 & 0 & 1  
    \end{pmatrix}
    \begin{pmatrix}
         x^{0}\\
         x^{1}\\
         x^{2}\\
         x^{3}
    \end{pmatrix}     
\end{equation}
where $\beta = v/c$.

As a matter of fact, \eqref{lorentz-transformation} induces the transformation of the components of any  $4-$vector, 
\begin{equation}\label{estrela}
    A^\mu \rightarrow A'^\mu = \Lambda^\mu{}_\nu A^\nu.
\end{equation}

With the transformation of vectors, let us see how $\mathcal{F}^\mu{}_\nu$ transforms under a Lorentz boost. Firstly, we may write, due to the principle of relativity, 
\begin{equation}\label{um-asterisco}
    f^\mu = q \mathcal{F}^\mu{}_\nu V^\nu \Leftrightarrow f'^\mu = q \mathcal{F}'^\mu{}_\nu V'^\nu,
\end{equation}
since the laws of physics hold in every (inertial) frame. The arrow points to both sides, after all, there are no privileged coordinates for different frames. According to \eqref{estrela}, the right part of \eqref{um-asterisco} reads,
\begin{equation}\label{dois-asterisco}
    \Lambda^\mu{}_\nu f^\nu = q \mathcal{F}'^\mu{}_\beta \Lambda^\beta{}_\alpha V^\alpha \Rightarrow f^\nu = q \left ( \Lambda^{-1} \right )^\nu{}_\mu \mathcal{F}'^\mu{}_\beta \Lambda^\beta{}_\alpha V^\alpha.
\end{equation}
Comparing \eqref{um-asterisco} and \eqref{dois-asterisco}, we conclude that
\begin{equation}\label{quadrado}
\mathcal{F}^\nu{}_\alpha = \left ( \Lambda^{-1} \right)^\nu{}_\mu \mathcal{F}'^\mu{}_\beta \Lambda^\beta{}_\alpha  \Leftrightarrow \mathcal{F}'^\mu{}_\beta  =   \Lambda^\mu{}_\nu \mathcal{F}^\nu{}_\alpha \left ( \Lambda^{-1} \right)^\alpha{}_\beta.   
\end{equation}
That is exactly the standard transformation law of a rank $(1,1)-$tensor, obtained upon the principle of relativity. We point out that defining tensors as objects that transform in a particular way might appear to be an \textit{ad hoc} assumption, made without a clear basis, potentially leading to confusion. Therefore, by invoking the principle of relativity, we can deduce such transformations, as illustrated in equation \eqref{quadrado}, without requiring any prior premises.

Let us consider an example that will be useful for our future discussions. Suppose a particular spatial region is permeated by a magnetic field pointing in the $x^2-$direction, $\Vec{B} = B_2 \hat{e}_2$. In this case, the electromagnetic field tensor has only two non-vanishing terms, $\mathcal{F}^1{}_3 = -\mathcal{F}^3{}_1 = -B_2$. How is the field perceived by a reference frame that is boosted in the $x^1$ direction, with a relative velocity of $\Vec{v} = v\hat{e}_1$? A direct calculation shows that
\begin{equation}\label{f'munu}
    \mathcal{F}'^\mu{}_\beta = \begin{pmatrix}
        0 & 0 & 0 & \beta \gamma B_2 \\
        0 & 0 & 0 & - \gamma B_2 \\
        0 & 0 & 0 & 0 \\
        \beta \gamma B_2 & \gamma B_2 & 0 & 0  
    \end{pmatrix}.
\end{equation}

However, the values in the first row (and column) of the matrix that represents the tensor should be interpreted as the electric field. Therefore, our previous result shows that the $\Vec{B}-$field in the initial frame is perceived as an electric field in the frame that is moving with respect to the former. Comparing the terms ${}^0{}_3$ in both \eqref{fmunu} and \eqref{f'munu}, we have
\begin{equation}\label{zero}
   E'_3 = v \gamma B_2.
\end{equation}

Once again, we emphasize that electric and magnetic fields are relative. An electric field can be perceived as a magnetic field and vice versa, in the sense of \eqref{zero}. Another interesting aspect of the transformation \eqref{zero} is its behavior in the limit where $v \ll c$: $E'_3 \approx v B_2$. Despite the low limit of speeds, we still observe a relative change from magnetic to electric fields. In contrast, a Galilean boost does not preserve the fundamental structure of $\mathcal{F}^\mu{}_\nu$, which includes a symmetric first row and column ($\mathcal{F}^0{}_i = \mathcal{F}^i{}_0$) in addition to the anti-symmetric sector $\mathcal{F}^i{}_j$.
\section{Brief recap on the induction law}
\label{sec4}

\begin{figure}[H]
    \centering
    \includegraphics[scale=0.3]{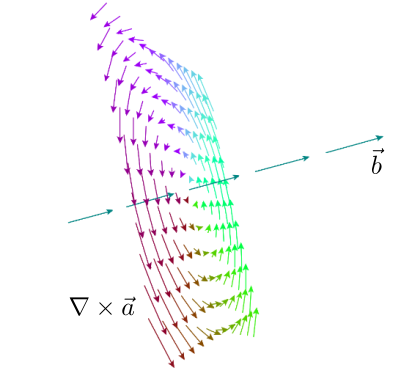} 
    \caption{Geometric picture of the equation \eqref{1.1}.}
    \label{fig:rot}
\end{figure}

In this section, we will discuss the very structure of the induction law through a simple experiment. Due to its geometrical appeal, we will focus on the differential form of the law. Prior to that, it is instructive to consider an equation of the form
\begin{equation}\label{1.1}
    \Vec{b} = \nabla \times \vec{a}.
\end{equation}
The geometrical meaning of Equation \eqref{1.1} is clear: if we interpret the vector field $\vec{b}$ as the source of the field $\vec{a}$, then the latter circulates around the former according to the right-hand rule, as shown in Figure \ref{fig:rot}.
In this particular case, $\vec{a}(x^1,x^2,x^3) = -x^2/2 \hat{e}_1 + x^1/2 \hat{e}_2$, which implies $\vec{b} = \hat{e}_3$. Here, once again, ${ \hat{e}_1, \hat{e}_2, \hat{e}_3 }$ are the unit vectors pointing in the $x^1$, $x^2$, and $x^3$ directions of a Cartesian coordinate system. There are many other examples with this composition, even within the realm of electromagnetism. For instance:
\begin{enumerate}
\item The Ampère's law indicates that currents generate a magnetic field around them, given by the equation $\nabla \times \vec{B} = \mu_0 \vec{J}$.
\item The magnetic field can be expressed as $\nabla \times \vec{A} = \Vec{B}$.
\end{enumerate}

\begin{figure}[H]
    \centering
    \includegraphics[scale=0.36]{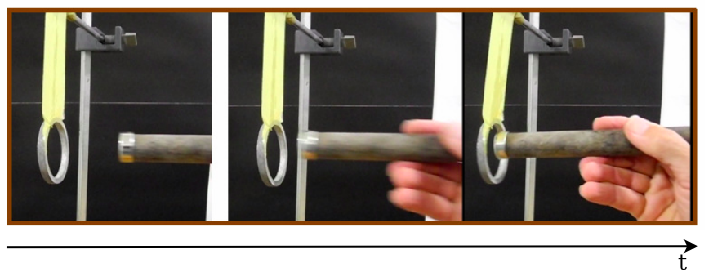} 
    \caption{The aluminum ring is repeled by the moving magnet. Source: \url{https://www.fisica.ufjf.br/~lesche/F\%C3\%ADs\%20III/Notas\%20de\%20aula\%20F\%20III/}}
    \label{fig:rep}
\end{figure}

The previous mathematical concepts will be directly applied to a simple yet astonishing experiment. We bring a magnet close to a conducting ring, and they are not supposed to interact (magnetically) with each other. In this case, the ring could be made of, say, silver or aluminum. However, a curious effect occurs when the magnet enters the circle defined by the ring: the system feels a repulsion, and the ring is deflected in an effort to move away from the magnet. This sequence is shown in Figure \ref{fig:rep}, where the line below it indicates the arrow of time.

Since the ring does not feel the presence of a static magnet, the only possible interaction between them must have been created by the variation of $\Vec{B}$ in time, given by $\partial \vec{B}/ \partial t$. Knowing that like magnetic poles repel each other, we expect that $\partial \vec{B}/ \partial t$ induced a current on the ring, denoted as $I_{ind}$, and the corresponding magnetic field created by the current ($\Vec{B}_{ind}$) interacted with the $\Vec{B}$ field of the magnet, causing the repulsion.
\begin{figure}[H]
    \centering
    \includegraphics[scale=0.26]{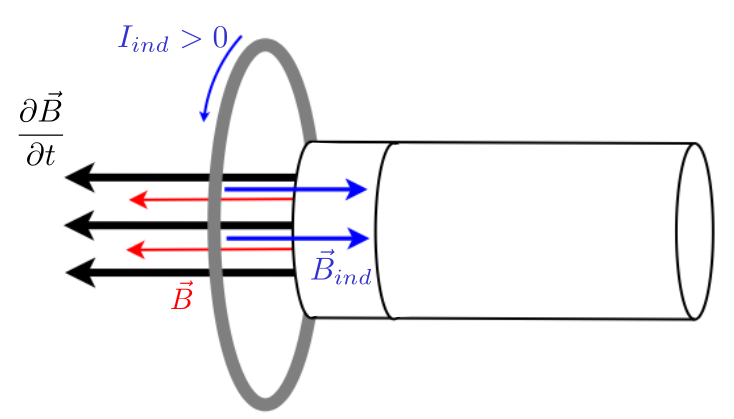} 
    \caption{Schematic representation of the repulsion between magnet and ring.}
    \label{fig:exp}
\end{figure}

And one might now ask, what is the origin of $I_{ind}$? Since the ring was initially stationary relative to the magnet, our experiment indicates that $I_{ind}$ was generated by an electric field $\vec{E}$ that can only have a specific configuration - it circulates around $\partial \Vec{B}/ \partial t$, obeying the \textit{left-hand} rule. Hence, our previous discussion, as seen in \eqref{1.1}, allows us to formulate the differential form of the induction law,
\begin{equation}\label{3.1}
    \frac{\partial \Vec{B}}{\partial t} = \lambda \nabla \times \Vec{E},
\end{equation}
here $\lambda$ is still to be determined.

First of all, a simple dimensional analysis shows that
\begin{equation}
[\lambda] = \frac{T/s}{N/mC} = 1,
\end{equation}
meaning that this parameter is a dimensionless constant. Due to the left-hand rule depicted in Figure \ref{fig:exp}, we expect it to be negative. 

To conclude this section, we would like to emphasize that the experiment can be repeated with the magnet polarization inverted, and it would work in the same way. Furthermore, if the magnet is moved away from the O-ring instead of being pushed towards it, the pair will experience an attraction, and the explanations provided above would still hold.

Next, we will investigate a particular experiment that allows us to obtain $\lambda$ by invoking the principle of relativity.

\section{Toy model for measuring currents}
\label{sec5}

\begin{figure}[H]
    \centering
    \includegraphics[scale=0.27]{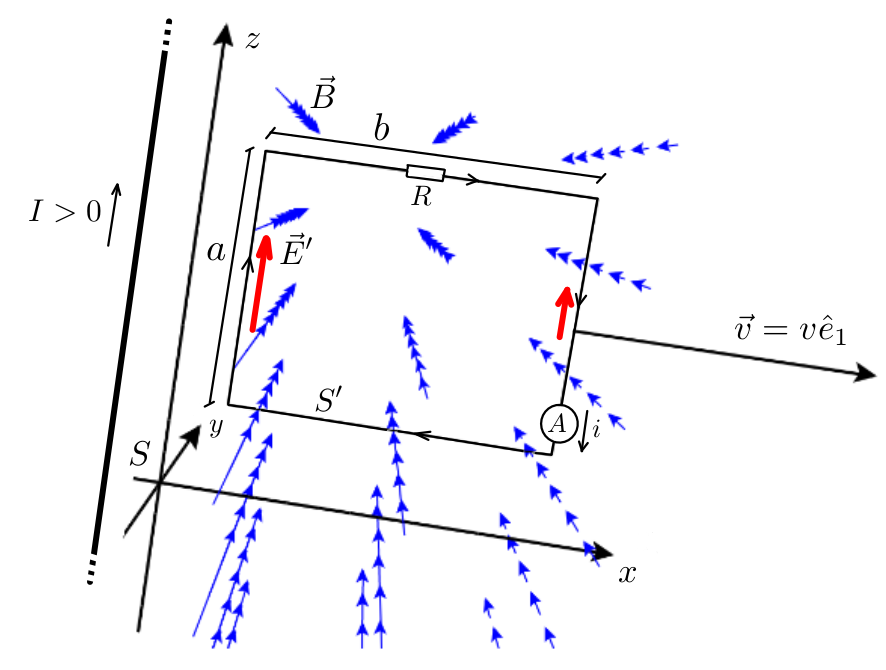} 
    \caption{Experimental setup used to explore the relativistic behavior of the induction law.}
    \label{fig:lenz}
\end{figure}

Let us consider a simple experiment consisting of an infinite wire carrying a steady current $I>0$, as indicated in Figure \ref{fig:lenz}. Here, we emulate a similar experiment presented previously with a different setup. At the same plane where the wire is located, there is a rectangular closed-loop circuit moving away from the wire with a fixed velocity $\Vec{v}$ orthogonal to the wire direction. The resistance of the circuit is given by $R$, and its sides are $a$ and $b$. An ammeter is connected to the circuit with the orientation indicated close to it. We point out that the arrow does not represent the physical effect of moving charges. Rather, it is only an experimental prescription of how to connect the ammeter in the circuit.

Thinking of the wire and the circuit as rigid bodies, they define reference frames that we name $S$ and $S'$, respectively. We fix a Cartesian coordinate system in $S$, in a way that the wire is in the $z$ direction, and $\Vec{v}$ points towards the positive $x$-direction.

An observer in $S$ may use the induction law \eqref{3.1} to estimate the current $i$. For that, we integrate \eqref{3.1} over the surface $(\mathcal{S})$ delimited by the closed path around the circuit, oriented clockwise, 
\begin{equation}\label{4.1}
    \iint\limits_{(\mathcal{S})} \frac{\partial \Vec{B}}{\partial t} \cdot \overrightarrow{dS} = \lambda \iint\limits_{(\mathcal{S})} (\nabla  \times \vec{E})\cdot  \overrightarrow{dS} = \lambda \oint\limits_{(\partial \mathcal{S})} \Vec{E} \cdot \overrightarrow{dl}.
\end{equation}
The last equality is justified by the Stokes' theorem and provides the voltage in the circuit induced in accordance with \eqref{3.1}.

Let us gather some pieces of information.

(\textit{i}) With the oriented ammeter, we have
\begin{equation}\label{5.1}
\oint\limits_{(\partial \mathcal{S})} \Vec{E} \cdot \overrightarrow{dl} = Ri.
\end{equation}

(\textit{ii}) The magnetic field generated by the wire is given by
\begin{equation}
\Vec{B} = \frac{\mu_0 I}{2 \pi x} \hat{e}_2.
\end{equation}

For this case, in which the wire is moving,  the flux in \eqref{4.1} may calculated by \cite{flanders_differentiation_1973, feoli_rotation_2021} 
\begin{align}\label{5.2}
    \iint \limits_{(\mathcal{S})} \frac{\partial \Vec{B}}{\partial t} \cdot \overrightarrow{dS} & = 
\frac{d}{dt} \iint \limits_{(\mathcal{S})} \Vec{B} \cdot \overrightarrow{dS} \cr & = \frac{d}{dt} \int^b_0 \int^{x+a}_a \frac{\mu_0 I}{2 \pi x}\hat{e}_2\cdot \hat{e}_2 dx dy \cr &= \frac{\mu_0 I b v}{2 \pi} \left ( \frac{1}{x+a} - \frac{1}{x} \right ),
\end{align}
where $dx/dt = v$ and $\Vec{dS} = dx dy \hat{j}$. We observe that both results \eqref{5.1} and \eqref{5.2} are related. Since $(\partial \mathcal{S})$ runs clockwise, $(\mathcal{S})$ is oriented in the $+ \hat{j}$ direction and vice versa. Finally, the current predicted by an observer in $S$ is given by
\begin{equation}\label{5.4}
    i= \frac{\mu_0 I b v}{2 R \pi \lambda} \left ( \frac{1}{x+a} - \frac{1}{x} \right ).
\end{equation}
On the other hand, an observer in $S'$ comoving with the circuit could also estimate the current. Our previous discussion on the relativity of electric and magnetic fields tells us that $\Vec{B}$ in $S$ is seen in $S'$ as an electric field $\Vec{E}'$, pointing toward the direction defined by $\Vec{v}\times \Vec{B}$. More precisely, the transformation of the electromagnetic strength tensor in \eqref{f'munu} provides
\begin{equation}\label{6.1}
    \Vec{E}'= \gamma \Vec{v}\times \vec{B} = \frac{\gamma v \mu_0 I}{2 \pi x}\hat{e}_3.
\end{equation}

Now, the current named $i'$, is given by
\begin{equation}\label{6.2}
    Ri'= \oint \limits_{(\partial \mathcal{S})}\Vec{E}'\cdot \overrightarrow{dl}' \Rightarrow i'= \gamma \frac{v\mu_0 I}{2\pi R}\left ( \frac{1}{x} - \frac{1}{x+a}\right ).
\end{equation}

Owing to the principle of relativity, we expect that there is only one possible value for the current. Thus, comparing \eqref{5.4} and \eqref{6.2} in the limit of low velocities ($\gamma \approx 1$), we can determine the value of $\lambda$ as
\begin{equation}
i=i' \Rightarrow \lambda = -1,
\end{equation}
leading to the well-known induction law
\begin{equation}
\nabla \times \Vec{E} = - \frac{\partial \Vec{B}}{\partial t}
\end{equation}
based on relativity arguments.

\section{Conclusions}
\label{sec6}
In this paper we have discussed the relativistic character of electromagnetism. Our findings are the following:
\begin{enumerate}
    \item From a operational approach, we have constructed the electromagnetic field tensor $\mathcal{F}^\mu{}_\nu$, which allows us to describe the Lorentz force when it acts on 4-velocity vectors.
    
    \item By utilizing the principles of special relativity, we have established the transformation rule that the tensor $\mathcal{F}^\mu{}_\nu$ must adhere to. This explicit demonstration highlights the inherently relativistic nature of electromagnetism. We recognize that what we perceive as electric and magnetic fields are merely different manifestations of the same underlying entity, contingent upon the reference frame from which we observe them.     
    
    \item Again, using an alternative operational approach, we successfully derived the induction law and demonstrated that the origins of Lenz's law are purely relativistic in nature.
\end{enumerate}

\section*{Acknowledgments}

This work has been supported by Programa Institucional
de Bolsas de Inicia\c{c}\~ao Cient\'ifica - 
XXXV BIC/Universidade Federal de Juiz
de Fora - 2022/2023, project number 51670.
\newline

\section*{ORCID iDs}

Thales B. S. F. Rodrigues  \url{https://orcid.org/0000-0002-1048-8586}\\
B. F. Rizutti \url{https://orcid.org/0000-0003-3667-1335}

\end{document}